# Absence of spontaneous time-reversal symmetry breaking and ferromagnetism in superconducting NiBi$_3$ single crystal

Jingyuan Wang,[1] Camron Farhang,[1] Di Yue,[2] Xiaofeng Jin,[2] Xiangde Zhu,[3] and Jing Xia[1, *]

[1]*Department of Physics and Astronomy, University of California, Irvine, California 92697, USA*
[2]*Department of Physics, Fudan University, Shanghai 200433, China*
[3]*Anhui Province Key Laboratory of Condensed Matter Physics at Extreme Conditions, High Magnetic Field Laboratory of the Chinese Academy of Sciences, Hefei 230031, Anhui, China.*

Recent experiments have pointed to a chiral p-wave-like superconductivity in epitaxial Bi/Ni bilayers that are spontaneously time-reversal symmetry breaking (TRSB), making it a promising platform for exploring physics useful for topologically protected quantum computing. Quite intriguingly, evidence has emerged that in non-epitaxial Bi/Ni bilayers, superconductivity arises due to the formation of NiBi$_3$, which has been reported to host coexisting ferromagnetic and superconducting orders at the surface. We perform high resolution surface magneto-optic Kerr effect (SMOKE) measurements using a Sagnac interferometer on single crystal NiBi$_3$ and find no sign of any spontaneous Kerr signal except for contributions from trapped vortices. This strongly indicates the absence of TRSB in NiBi$_3$, whether due to TRSB in the superconducting state or any coexisting ferromagnetism, and we conclude that the superconductivity found in non-epitaxial Bi/Ni is distinctively different from that in epitaxial Bi/Ni.

The quest to build a reliable quantum computer has stimulated intense research into quantum phases with quasiparticles that obey non-Abelian exchange rules and can be used for topologically protected quantum computing [1]. Such quasiparticles would exist as Majorana bound states in the vortex cores of a chiral p-wave superconductor [1,2], which is an electronic analog to the A-phase of superfluid $He^3$ [3] and breaks time-reversal symmetry (TRS). In the prototypical chiral p-wave superconductor Sr$_2$RuO$_4$ ( $T_C \approx$ 1.5 $K$), although TRS-breaking (TRSB) has been confirmed by Muon Spin Relaxation ($\mu SR$) [4], surface magneto-optic Kerr effect (SMOKE) measurements using a Sagnac interferometer [5], and $\mu SR$ under strain [6], the p-wave aspect has been challenged by the recent nuclear magnetic resonance (NMR) evidence [7] for an even-parity superconducting order parameter. In addition, a magnetic competing order has been identified in close proximity [8] by $\mu SR$ [6] and elastocaloric effects [8], making the picture of Sr$_2$RuO$_4$ rather complicated.

Superconducting epitaxial Bi/Ni bilayers provide a promising alternative candidate for chiral p-wave superconductivity. It was initially found in tunneling measurements that Bi layers deposited on Ni layers become superconducting with $T_C \approx 4\ K$ [9], and there are coexisting superconducting and ferromagnetic gaps when tunneling from the Ni side [10]. More recently, in high quality Bi/Ni bilayers grown by molecular beam epitaxy (MBE), superconducting quantum interference device (SQUID) measurements [11] show evidence for chiral superconductivity and the formation of chiral domains. SMOKE measurements using a Sagnac interferometer [12] conducted on the Bi side reveal spontaneous TRSB in the superconducting state, where chirality can be trained by a small magnetic field $\sim 100\ Oe$. Assuming that superconductivity exists only in the top Bi surface away from Ni, we have proposed a $d_{xy} \pm id_{x^2-y^2}$ superconducting order parameter, which is the lowest angular momentum state allowed by this surface symmetry [12]. This hypothetical restriction was soon corrected by a Time-domain Terahertz (THz) spectroscopy experiment [13] that identified a nodeless superconductivity extending over the entire Bi/Ni bilayer. Their data also rule out the odd-frequency pairing [14], which is natural for a superconductor-ferromagnet interface. These experimental findings collectively point to chiral p-wave superconductivity in strongly spin-orbit coupled epitaxial Bi/Ni bilayers [15], whose properties can in principle be engineered by the growth parameters (thickness, strain, doping) to optimize the conditions for hosting Majorana particles.

Real materials are complex. A radically different picture has emerged in Bi/Ni bilayers fabricated using other methods, highlighting the role of the intermetallic compound NiBi$_3$. NiBi$_3$ impurities were first detected in thermally evaporated Bi/Ni bilayers by X-ray diffraction (XRD) [16] and were proposed as the source for the observed superconductivity. Later studies on pulse-laser deposited (PLD) [17] and sputter deposited [18] Bi/Ni bilayers show the absence of superconductivity in as-grown samples without NiBi$_3$ impurities. By changing the deposition temperature [17], or by weeks of annealing [18], these samples develop superconductivity coincident with the formation of NiBi$_3$. As a known type-II s-wave superconductor with $T_C \approx 4\ K$ [19,20], NiBi$_3$ should be TRS-invariant, but there are reports of coexisting ferromagnetism and superconductivity in NiBi$_3$. Extrinsic ferromagnetism was found in flux-grown NiBi$_3$ crystals due to amorphous Ni impurities [21]. Intrinsic magnetic orders were proposed at the surface due to modifications of surface electronic band structures [22]: SQUID magnetometry has identified ferromagnetism in NiBi$_3$ nano-strains (200 nm) with high surface fraction [22]; electron spin resonance

(ESR) has detected no ferromagnetism but found surface induced magnetic fluctuations in single crystal NiBi$_3$ [23].

Although these reports of magnetic orders in NiBi$_3$ differ quantitatively from the TRSB observed in epitaxial Bi/Ni bilayer by Sagnac interferometry [12], and the coexistence of ferromagnetism and superconductivity often leads to odd-frequency pairing [14] that is inconsistent with THz time-domain spectroscopy data [13], it is sometimes argued that the observed unconventional superconductivity in epitaxial Bi/Ni bilayer may come from superconducting NiBi$_3$ impurities that have surface-induced ferromagnetism. Does NiBi$_3$ break TRS? Is it ferromagnetic near the surface? Above all, do epitaxial and non-epitaxial Bi/Ni bilayers host identical or distinct superconducting states? These fundamental questions can be addressed by performing a definitive determination of the TRS and magnetic properties of single crystal NiBi$_3$, especially near the surface.

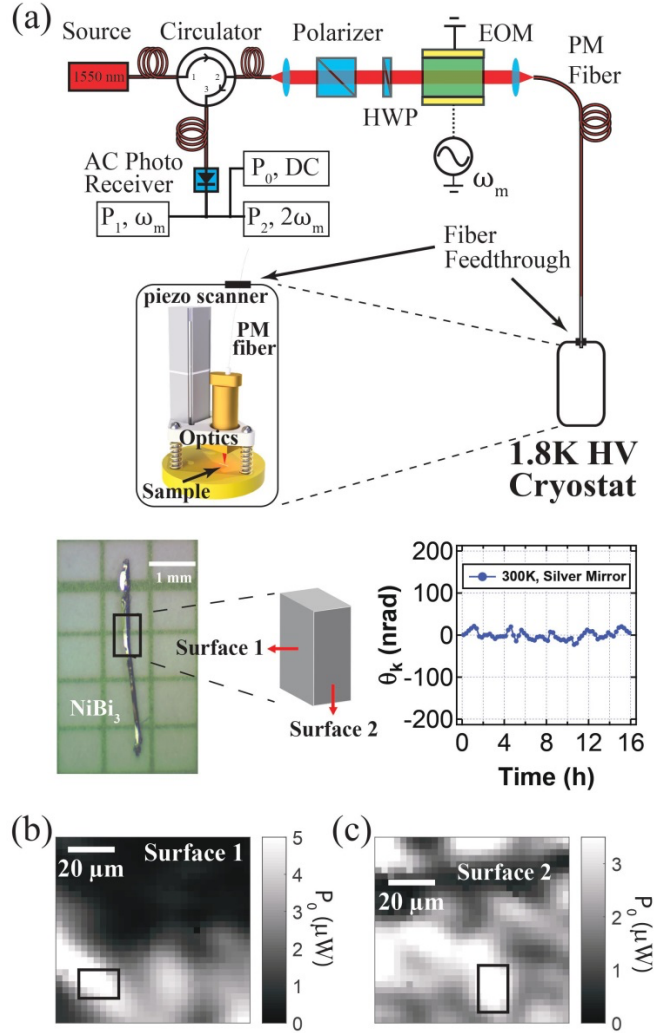

at 1.8 $K$ on surfaces 1 and 2, with black boxes marking optically flat regions for measurements.

SMOKE [24,25] measurements performed by a zero-area loop fiber optic Sagnac interferometer [26] are ideally suited for performing such a definitive test of TRSB and ferromagnetism near the surface of NiBi$_3$. Probing the sample surface with an optical penetration depth $\delta$ that is typically a few nanometers for conductors [24,25], SMOKE has proven to be a powerful probe for surface magnetization. Primarily for detecting even smaller Kerr signals that arise in unconventional superconductors, we have introduced a zero-area loop [26] fiber optic Sagnac interferometer that measures directly the non-reciprocal phase difference $\theta_{nr} = 2\theta_K$ between counter-propagating circularly polarized light beams, where $\theta_K$ is the Kerr rotation. This approach fundamentally rejects polarization rotations due to non-TRSB effects such as linear and circular dichroism [28]. This design has pushed the Kerr resolution from microradian ($\mu rad$) [24,25] to ten nanoradian ($nrad$) level [5], allowing us to identify TRSB in various unconventional superconductors such as Sr$_2$RuO$_4$ [5] and Bi/Ni bilayers [12]. Scanning imaging capability with $\mu m$ spatial resolution has allowed us to discover ferromagnetism in 2D van der Waals layers [29] and to control magnetism in 2D structures [30]. We use a scanning Sagnac microscope operating at 1550 nm wavelength as illustrated in Fig. 1(a). The interferometer itself is located at room temperature. And the piezo scanner [31] is mounted inside a cryostat with 1.8 $K$ base temperature and 9 $T$ magnetic field capability. A polarization maintaining fiber delivers lights of orthogonal linear polarizations into the high vacuum sample space inside the cryostat. And a cryogenic quarter wave ($\lambda/4$) plate converts these light beams into circular polarizations of opposite chiralities that will interact with the sample surface and detect TRSB. Fig. 1(b) shows a 16-hour measurement on a silver mirror demonstrating 10 $nrad$ Kerr resolution that is limited by long-term drifts in optics and electronics.

FIG. 1. **Sagnac interferometer and NiBi$_3$ crystal** (a) Schematics of a scanning Sagnac microscope at 1550 nm wavelength (top), NiBi$_3$ crystal (left) and 16-hour Sagnac drift test on a silver mirror showing 10 $nrad$ Kerr resolution (right). (b), (c) Reflected optical power ($P_0$) map

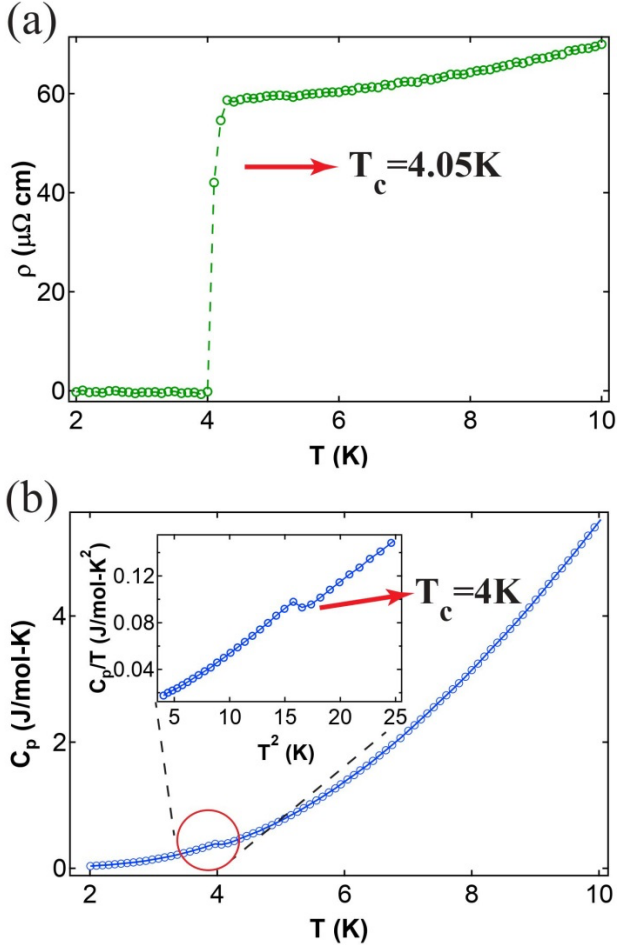

FIG. 2. **Resistivity and specific heat** (a) Resistivity ($\rho$) of NiBi$_3$, where $T_C \sim 4\ K$ is determined as the middle point of the resistivity drop. (b) Specific heat ($C_p$) with a kink at $T_C$. The inset shows $C_p/T$ vs. $T^2$ near $T_C$.

Needle-shaped NiBi$_3$ single crystals were grown using the self-flux method with the $b$ axis along the longest dimension (Fig. 1(a)) as determined by x-ray diffractometry [23]. The typical size of such a single crystal is $\sim 3\ mm\ \times\ 0.2\ mm\ \times\ 0.2\ mm$. Fig. 2(a) shows the measured resistivity $\rho$ of the NiBi$_3$ sample near the superconducting transition, with the excitation current flowing along the $b$ axis. $T_C = 4.05\ K$ is determined as the middle point of the resistivity drop, and is in good agreement with the result in ref [23] on the same batch of crystals. The specific heat ($C_p$) is shown in Fig. 2(b), with $C_p/T$ vs. $T^2$ plotted near $T_C$ in the inset. A prominent kink at $\sim 4\ K$ indicates a sudden change in the Fermionic contributions to $C_p$, and confirms the superconducting transition. We note that anomalies in $C_p$ around 2.2 $K$ have been reported [21] in NiBi$_3$ due to amorphous Ni impurities, but we observe no such anomaly in our $C_p$ data, attesting to the high quality of crystals used in this study.

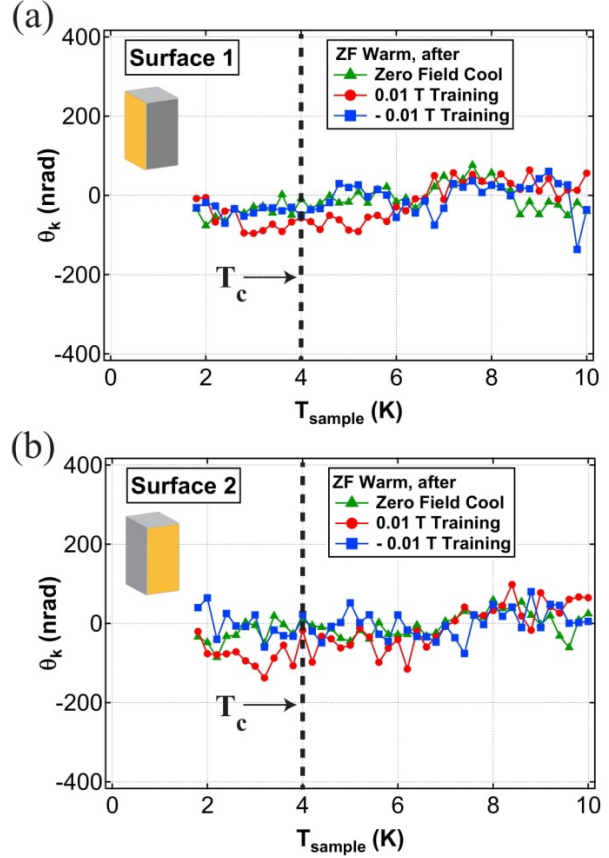

FIG. 3. **Absence of TRSB in the superconducting state.** Kerr signals measured on (a) surface 1 and (b) surface 2 during zero-field (ZF) warmups, after zero field cooldown or after $\pm$ 0.01 T field "trainings", showing no TRSB.

SMOKE measurements are performed on two lateral surfaces of the crystal, dubbed surface 1 and surface 2 that are perpendicular to the $a$ and $c$ axis, as shown in Fig. 1(a). Due to the softness of the crystal, the surfaces of as-grown crystals are curved. It is necessary to perform low-temperature scanning imaging to locate optically flat regions for SMOKE measurements. Fig. 1(b) and (c) are images of reflected light power ($P_0$) from surface 1 and surface 2 respectively, and optically flat regions marked by black boxes are chosen for SMOKE measurements with $P_0 \sim 5\ \mu W$.

To test possible spontaneous TRSB in the superconducting state, we perform SMOKE measurements at fixed locations on surfaces 1 and 2 during zero-magnetic-field (ZF) warmups. Kerr signals $\theta_K$ of such ZF warmups after zero-field cooling are presented as green curves in Fig. 3(a)(b), showing no sign of TRSB with an uncertainty of 20 $nrad$ across $T_C$. As is typical of spontaneous TRSB, the sign and size of $\theta_K$ at zero magnetic fields normally vary as a function of location and temperature. Therefore, a small training field $B_{training}$ is often applied and then removed to align the chiral domains in SMOKE measurements of unconventional superconductors such as in the studies of $Sr_2RuO_4$ [5], $UPt_3$ [32], $UTe_2$ [33] to name a few. It is noted that in all these examples, $B_{training}$ is chosen to be smaller than the lower critical field $H_{C1}$ to avoid introducing vortices that can be trapped at pinning sites even after the removal of the training fields. Trainings with $B_{training} > H_{C1}$ could result in non-zero $\theta_K$ during ZF warmups due to contributions from trapped vortices, such as those found in $YBa_2Cu_3O_{6+x}$ with a $4\,T$ training field [34]. We pick $B_{training} = \pm 0.01$ T for $NiBi_3$, which is smaller than the measured value [20] of $H_{C1} = 0.015$ T. Kerr signal $\theta_K$ during ZF warmups after $\pm 0.01$ T trainings are plotted as red and blue curves in Fig. 3(a)(b) for surfaces 1 and 2 respectively: no spontaneous $\theta_K$ is observed across $T_C$ with an uncertainty of $20\,nrad$. In comparison, in epitaxial Bi/Ni bilayers of $20\,nm$ thickness [12], we have detected $\theta_K \sim 120\,nrad$ onsetting abruptly at $T_C = 4.1\,K$ [12]. We can therefore conclude that there is no sign of spontaneous TRSB in the superconducting state of single crystal $NiBi_3$. Furthermore, it was found in sputtered Bi/Ni bilayers, the $NiBi_3$ impurity phase has a preferred orientation of (203) [18]. This translates to a crystalline surface parallel to the $b$ axis, which corresponds to either surface 1 or surface 2 that is measured here. Therefore, we could rule out TRSB superconductivity in sputtered and PLD Bi/Ni bilayers where $NiBi_3$ is responsible for superconductivity [17,18].

Now we turn to tests of possible ferromagnetism in $NiBi_3$ that could be induced by either surface effects [22] or extrinsic Ni impurities [21]. As explained earlier, heat capacity $C_p$ (Fig. 2(b)) in our samples indicates a much lower impurity level compared to those used in ref [21], and unlike bulk SQUID magnetometry, Sagnac probes an optical volume of only $\sim 0.1\,\mu m^3$, making it much less susceptible to Ni impurities.

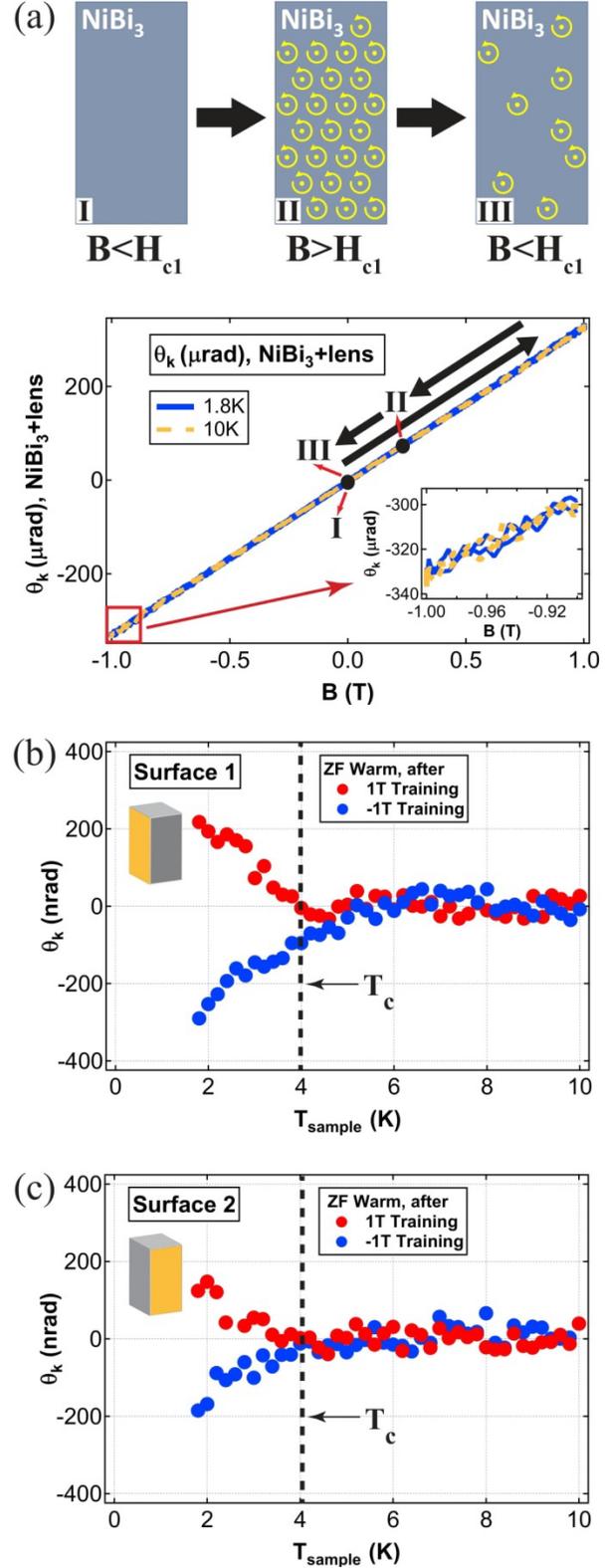

FIG. 4. **Trapped vortices and absence of ferromagnetism.** (a) Illustration of trapped vortices after removal of a magnetic field $> H_{c1}$ (top); Kerr signals during $1\,T$ magnetic field hysteresis on Surface 2 at $1.8\,K$ and $10\,K$ (bottom). (b)

and (c) are Kerr signals measured during zero-field warmups after removing $\pm 1\,T$ field on surface 1 and surface 2 respectively, showing $\theta_K \sim \pm 200\,nrad$ onsetting at $T_C$ due to trapped vortices. There is no sign of any ferromagnetism.

We first perform magnetic hysteresis measurements with magnetic fields up to $\pm 1\,T$, which is similar to the conditions in ref [22]. These are shown in Fig. 4(a) for $T = 1.8\,K < T_C$ (blue) and $T = 10\,K > T_C$ (yellow). The Kerr signals are extremely linear with the magnetic field $B$. They are dominated by the background Faraday effect contribution from the low-temperature objective lens, which is proportional to $B$. The higher noise level $\Delta\theta_K$ comes from the fluctuations in the above lens contribution induced by magnetic field noise. $\Delta\theta_K \sim 5\,\mu rad$ at high magnetic fields can be seen in the inset of Fig. 4(a) for $\theta_K$ taken between $B = -1\,T$ and $-0.9\,T$. Unlike in ref [21,22], we observe no sign of any ferromagnetic hysteresis with $5\,\mu rad$ uncertainty. It is worth noting that using the same instrument we have measured $\theta_K \sim 130\,\mu rad$ in $2\,nm$ of Ni [12], and $\theta_K \sim 500\,\mu rad$ in $4\,nm$ of SrRuO3 [30]. Therefore, this is already a strong constraint on any ferromagnetism in NiBi3.

For an even more stringent test of ferromagnetism, we measure the remanent Kerr signal by reducing the $1\,T$ magnetic field back to zero at $T = 1.8\,K$, as shown in the sequence I-II-III in Fig. 4(a). NiBi3 is a type-II superconductor with a lower critical field $H_{C1} = 0.015$ [20] and an upper critical field $H_{C2} = 0.35\,T$ [20]. As illustrated in the cartoon in Fig. 4(a), when $H_{C1} < B < H_{C2}$, vortices penetrate the superconducting sample. Their contributions to $\theta_K$ are linear with the magnetic field but are overwhelmed in the hysteresis measurements (Fig. 4(a)) by the much larger Faraday effect of the objective lens. After the magnetic field is removed (step III), a small fraction of vortices can be trapped at pinning sites, and they will contribute to $\theta_K$ during subsequent ZF warmups. The trapped vortices' contribution to $\theta_K$ would decrease exponentially as the temperature is raised towards $T_C$. The remanent Kerr signals during ZF warmups after $\pm 1\,T$ trainings are plotted in Fig. 4(b) for surface 1 and in Fig. 4(c) for surface 2 respectively. There are clear remanent Kerr signals of $\theta_K \sim \pm 200\,nrad$ onsetting sharply at $T_C$ due to trapped vortices. However, we observe no sign of any ferromagnetism with $20\,nrad$ uncertainty, unless its Curie temperature coincides precisely with $T_C$, which is highly unlikely. We note that the $20\,nrad$ uncertainty is four orders of magnitude smaller than the measured $\theta_K$ values in $2\,nm$ of Ni [12] or $4\,nm$ of SrRuO3 [30], strongly indicating that ferromagnetism is absent in NiBi3. Therefore, the reported ferromagnetism in nano-strains [22] of NiBi3 is not due to the surface of NiBi3, but must originate from other sources that are likely irrelevant to Bi/Ni bilayers.

In summary, we have provided strong error bounds of $20\,nrad$ for any spontaneous Kerr signals in single crystal NiBi3, strongly indicating the absence of TRSB in NiBi3, whether due to the superconducting state or any coexisting ferromagnetism. We can therefore conclude that the superconducting phases in epitaxial and non-epitaxial Bi/Ni bilayers are distinctively different. In non-epitaxial Bi/Ni, superconductivity originates from the formation of an impurity NiBi3 phase [17,18], which doesn't host coexisting ferromagnetic order or TRSB superconductivity. In contrast, the epitaxial Bi/Ni samples such as those grown by MBE host a superconducting state that is most likely to be of chiral p-wave based on existing experimental evidence [12,13,35]. The latter can be a promising platform for hosting Majorana particles useful for topologically protected quantum computing. And it is important to refine the growth process [17] to enable epitaxial growth, especially for non-MBE growth methods, to stabilize and optimize the chiral p-wave state for exploring Majorana physics for robust quantum computing applications.

Experiments at UC Irvine were supported by NSF award DMR-1807817, and in part by the Gordon and Betty Moore Foundation through Grant GBMF10276 to J.X.. The works at Hefei was supported by the Youth Innovation Promotion Association of CAS (Grant No. 2021117).